\documentstyle[11pt,newpasp,twoside,epsf]{article}
\def\edcomment#1{\iffalse\marginpar{\raggedright\sl#1\/}\else\relax\fi}
\reversemarginpar

\begin{document}
\title{Searching for Clusters of Galaxies with SUMSS}
\author{H. J. Buttery, G. Cotter}
\affil{Astrophysics, Cavendish Laboratory,
 Cambridge, CB3 0HE, UK.}
\author{R. W. Hunstead, E. M. Sadler}
\affil{School of Physics, University of Sydney,
      NSW 2006,
      Australia.}

\begin{abstract}
Statistical overdensities of radio sources in the NRAO VLA Sky
Survey (NVSS)
catalogue have proven to be signposts to high-redshift clusters of galaxies. A
similar search for overdensities has been carried out in the Sydney University Molonglo Sky
Survey (SUMSS), which is closely matched in resolution and frequency
to the NVSS. Sixty potential southern-hemisphere clusters have been found
in SUMSS.

\end{abstract}

\section{Introduction}
Large numbers of
distant ($z > 0.3$) clusters are now being identified by the new
generation of X-ray telescopes, such as Chandra and XMM-Newton
(Allen~et~al. 2001; 
Fabian~et~al. 2001). This will allow the study of the evolution of
gas and galaxies 
within massive dark-matter halos. However, because of the $n^2$ dependence of thermal bremsstrahlung, X-ray
selection of clusters is inevitably biased towards relaxed systems. 
Other cluster selection
techniques, 
which can also reach high redshifts, are needed to complement the X-ray work.

A different approach to cluster
selection is to look for
overdensities of radio sources. Such overdensities are known to
be signposts for high-redshift clusters of galaxies containing several
radio-loud AGN. 
These clusters will form a very different sample to that of the
high-redshift X-ray clusters; 
plausibly it will be biased towards merging systems
if, for example, AGN activity is triggered by mergers.

This search technique has been carried out on the
NRAO VLA Sky Survey (NVSS) with a high success rate.  Optical observations 
indicate that approximately $65\%$ of the overdensities of
radio sources are indeed 
clusters of galaxies (Croft \& Rawlings, private communication). 
Many $z> 0.3$ clusters have been found; TOC J0233.3+3021, which has optical and infrared
magnitudes implying $z\sim 1$,
 was mapped by the Ryle Telescope at
Cambridge and was found to have  a detectable Sunyaev-Zel''dovich
effect (Cotter et al. 2001).

\section{Continuing the search in the Southern Hemisphere}

The Sydney University Molonglo Sky Survey (SUMSS) is comparable to the
NVSS survey. It has a resolution of 43 arcseconds at 843 MHz and
identifies sources down to a level of 5 mJy.%, whereas the NVSS %
%has a resolution of 45 arcseconds at 1400 MHz and identifies sources
%down to a level of 2.5 mJy.% 
~It follows that overdensities in this survey will also be signposts to clusters of galaxies. 

We searched the SUMSS catalogue (Mauch, private
communication) for overdensities of 5 or more
radio sources inside a 7-arcminute radius.
To weed out
nearby clusters, we examine the SuperCOSMOS $R$-band data (Hambly et
al. 2001). About $5\%$
 of the radio source overdensities prove to be Abell clusters,
and another $35\%$ have one or more optical identifications to the SuperCOSMOS
limit of $R = 21.5$. The remaining overdensities are robust candidate
$z > 0.3$ galaxy clusters. Sixty such candidates have been found to
date; two are shown in Figure 1.

We will now embark on a detailed follow-up program that includes
higher-resolution radio imaging at the ATCA, deeper imaging at the
MSSSO 2.3-m and wide-field NIR imaging at the AAT.
When we have assembled a sample
of spectroscopically-confirmed clusters (using 8-m class telescopes for 
the most distant clusters), comparisons will be drawn
 between high-redshift X-ray- and radio source-selected samples. This
will allow examination, 
for example, of the hypothesis that merging  triggers multiple AGN
activity in clusters.

\begin{figure}
\plottwo{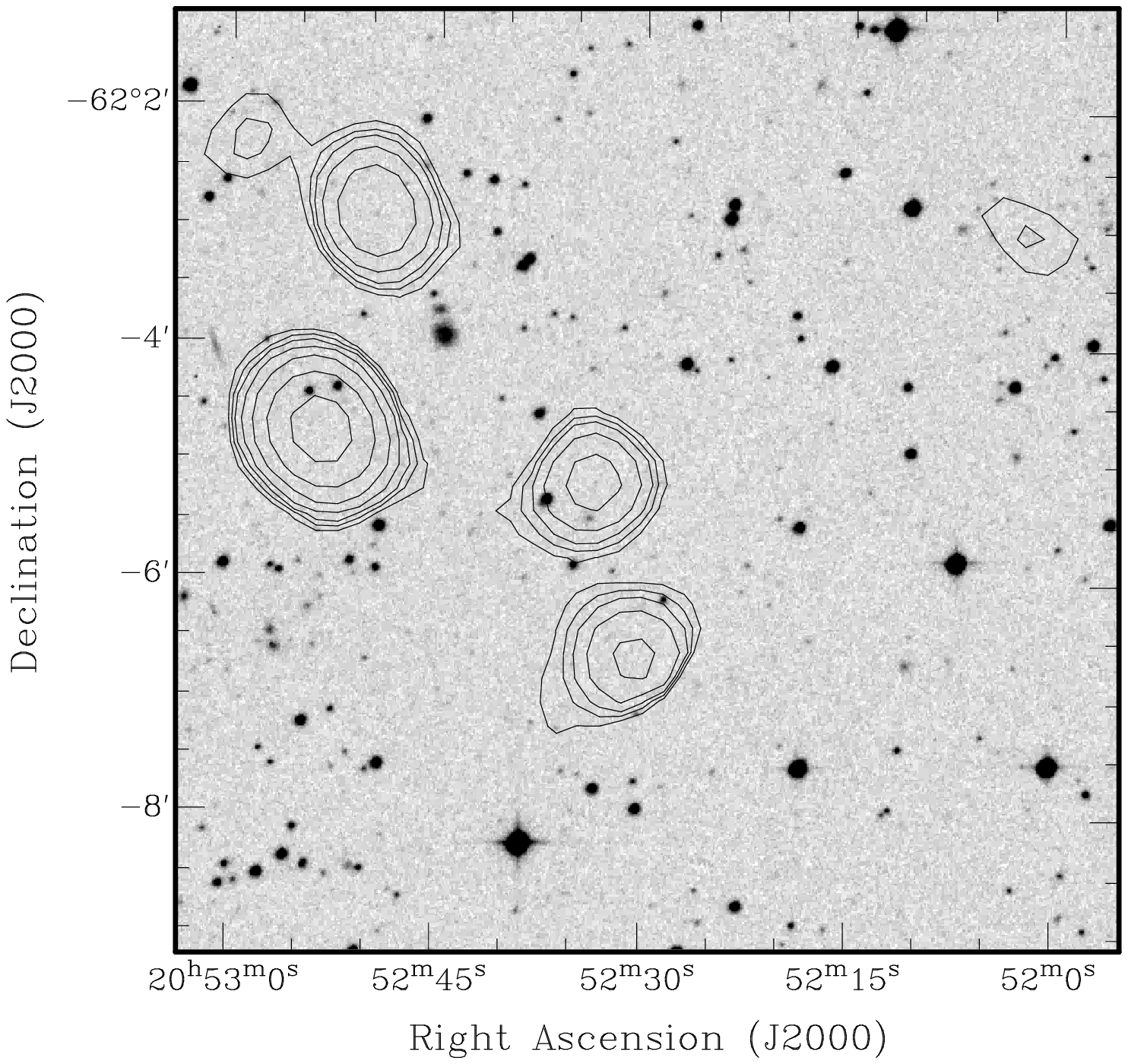}{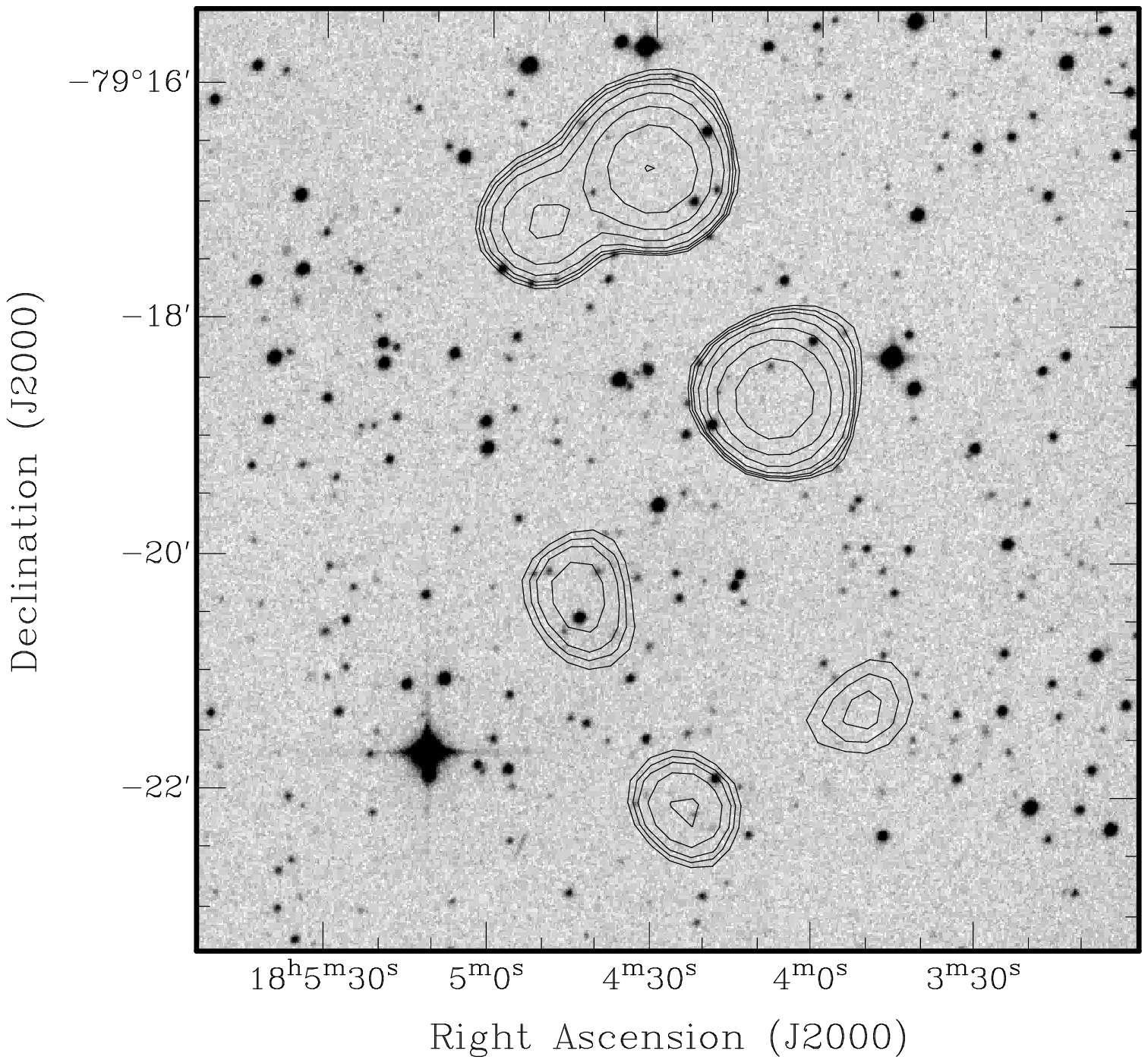}
\caption{Two examples of clusters found in the SUMSS
catalogue. The contours are from SUMSS and are at 4, 5, 6, 8, 12, 20, 36, 68
and 120 mJy/beam. The greyscale is an $R$-band image from
SuperCOSMOS. Few of the sources have optical identifications,
indicating that if these are clusters they will be at $z > 0.3$.}
\end{figure}
\acknowledgments 
HJB acknowledges a PPARC PhD studentship. GC acknowledges a PPARC
Postdoctoral Research Fellowship.

\end{document}